\documentclass[runningheads]{llncs}
\usepackage[belowskip=-15pt,aboveskip=0pt]{caption}
\usepackage{graphicx}
\usepackage[table,xcdraw]{xcolor}
\usepackage{graphicx}
\usepackage{amsmath}
\usepackage{float}
\usepackage{bm}
\usepackage{bbm}

\usepackage{subfig}

\usepackage[utf8]{inputenc} 
\usepackage[T1]{fontenc}    
\usepackage{hyperref}       
\usepackage{url}            
\usepackage{booktabs}       
\usepackage{amsfonts}       
\usepackage{nicefrac}       
\usepackage{microtype}      

\begin{document}
\title{Leveraging Uncertainty in Deep Learning for Pancreatic Adenocarcinoma Grading 
}
\titlerunning{AI in Pancreatic Adenocarcinoma Grading}
%
\author{Biraja Ghoshal \inst{1}
Bhargab Ghoshal\inst{2} \and
Allan Tucker \inst{1}}

%
%
\institute{Department of Computer Science, Brunel University, London, UK \and
Faculty of Medical Sciences, University College London, UK
\email{Biraja.Ghoshal@brunel.ac.uk}\\
}
\maketitle              
\begin{abstract}

Pancreatic cancers have one of the worst prognoses compared to other cancers, as they are diagnosed when cancer has progressed towards its latter stages. The current manual histological grading for diagnosing pancreatic adenocarcinomas is time-consuming and often results in misdiagnosis. In digital pathology, AI-based cancer grading must be extremely accurate in prediction and uncertainty quantification to improve reliability and explainability and are essential for gaining clinicians’ trust in the technology. We present Bayesian Convolutional Neural Networks for automated pancreatic cancer grading from MGG and H\&E stained images to estimate uncertainty in model prediction. We show that the estimated uncertainty correlates with prediction error. Specifically, it is useful in setting the acceptance threshold using a metric that weighs classification accuracy-reject trade-off and misclassification cost controlled by hyperparameters and can be employed in clinical settings.

\end{abstract}
\section{Introduction}

Pancreatic adenocarcinoma has one of the worst prognoses compared to other cancers, with a 10\% 5 year survival rate, and it is projected to become the second-leading cause of cancer-related mortality by 2030 \cite{park2021pancreatic}. Early diagnosis is important to improving the likelihood of survival, which is reflected in the survival rates of 39.4\% in patients whose tumours have not metastasised \cite{kenner2021artificial}. Intraepithelial neoplasms develop into adenocarcinomas and these are stratified into three progressive categories: Grade I, II and III. Grade I neoplasia comprises of well-differentiated tissue, which do not divide rapidly and so do not respond very well to chemotherapy. Grade II neoplasia represents moderately differentiated tissue, whilst Grade III represents poorly differentiated tissue, which is rapidly proliferating. The grading of neoplasia can be performed using histological samples stained using Haemotxylin and Eosin (H\&E) and May-Grunwald-Giemsa stain (MGG). H\&E stains nuclear components blue-purple and cytoplasmic components pink, whilst MGG stains help to ascertain the morphology of cells. 

Pancreatic cancers are usually detected towards their latter stages in locally advanced (30\%-35\%) or metastatic (50\%-55\%) stage as most patients are asymptomatic in the early stages \cite{hidalgo2010pancreatic}. As a result, pancreatic tumour diagnosis requires urgent action and a definite surgical plan. Pancreatic cancer detection and classification has either focused only on distinguishing benign tumor from malignant ones or tumor presence on radiology images \cite{naito2021deep}. The diagnostic performance by means of manual examination of pancreatic cancer grading is very tedious, time consuming, depends on clinicians' experience, and often results in misdiagnosis and thus incorrect treatment. There is an urgent need for novel methods to supplement radiologist interpretations in improving the sensitivity of pancreatic cancer detection from histopathology images.

Several methods use deep learning in cancer detection and diagnosis such as the Gleason grading of prostate cancer, colon cancer grading, and breast cancer detection \cite{anaya2021overview,rajpurkar2022ai,esteva2019guide,litjens2017survey}. However, it is critical to estimate uncertainty in medical image analysis as an additional insight to point predictions to improve the reliability in making decisions. Our objective is not to achieve state-of-the-art performance, but rather to evaluate the usefulness of estimating uncertainty approximating Bayesian Convolutional Neural Networks (BCNN) with Dropweights to improve the diagnosis.

In digital pathology, multi-gigapixel Whole Slide digitised histopathology Images (WSI) are divided into small patches because (i) different chemical preparations typically render different slides for the same piece of tissue, (ii) they are generated by different digitisation devices, (iii) the device settings may produce different images from the same slide and (iv) size limitations for CNN image inputs \cite{esteva2021deep}.
 
It is crucial to accurately grade pancreatic cancer in patients where the cost of an error is very high. In order to avoid misdiagnoses, it is necessary to estimate uncertainty in a model's predictions for automatically handling clear-cut diagnoses, whilst elevating difficult decisions to medical professionals, who can request further scans, recognizing the uncertainty and seek assistance.

There are many methods proposed for quantifying uncertainty or confidence estimates in deep learning such as Hamiltonian Monte Carlo, Stochastic Gradient Langevin Dynamic (SGLD), Laplace approximation, Bayes by backprop,  Deep ensembles, Monte Carlo (MC) dropout, MC-dropweight and MC-batch normalization \cite{lakshminarayanan2017simple,Gal_Thesis,Neal,Blundell,MacKay,ghoshal2021estimating_ci,Kendall,ghoshal2021bayesianAL}. There are also many measures to estimate uncertainty such as softmax variance, predictive entropy, mutual information, BALD \cite{houlsby2011bayesian}, Uncertainty measure proposed by Leibig \cite{leibig2017leveraging} and Feinman \cite{feinman2017detecting} and averaging predictions over multiple models, which are mostly focused on rejection accuracy and log-likelihood without assessing the quality of predictive uncertainty with calibrated expectations. 

In this paper, we propose Bayesian Convolutional Neural Networks (BCNN) approach to predict pancreatic cancer grading from two different May-Grunwald-Giemsa (MGG) and Haematoxylin and Eosin (H\&E) stained images and show that the estimated uncertainty in prediction has a strong correlation with classification accuracy, thus enabling the identification of false predictions or unknown cases. We present the first approach (to the best of our knowledge) in leveraging uncertainty in automated grading of pancreatic cancer based on histopathology images. The proposed Bayesian deep learing model can be very useful to clinicians in diagonising cancer grading system, which can address the problems in manual grading. We believe that the availability of uncertainty-aware deep learning solution will enable a wider adoption of Artificial Intelligence (AI) in a clinical setting.

\section{Proposed Method}

\subsection{Dataset}

Our approach was based on the patch-level prediction to include all tumor histological subtypes, ensuring that the selected patches were widely representative for practical diagnosis in order to be adaptable to meet the input size of most neural networks training and derive the likelihood of neoplasia at patch-level. The patches of approximately 200 X 200 pixels of non-overlapping regions in a WSI were firstly extracted from 138 high-resolution tissue-samples stained with MGG and H\&E \cite{sehmi2021pancreatic} with varying dimensions annotated with Normal, Grade-I, Grade-II and Grade-III. Overall, 49.5\% (3201) of the patches were selected after discarding patches with non-tissue information. The flow chart is shown in Fig. 1.  An imbalanced number of patch images in each class i.e. bias towards the majority proportions of cancer cells is reduced by class weight to regularise the loss function.

\begin{figure}[hp]
  \centering
  \subfloat {\includegraphics[width=1.0\textwidth]{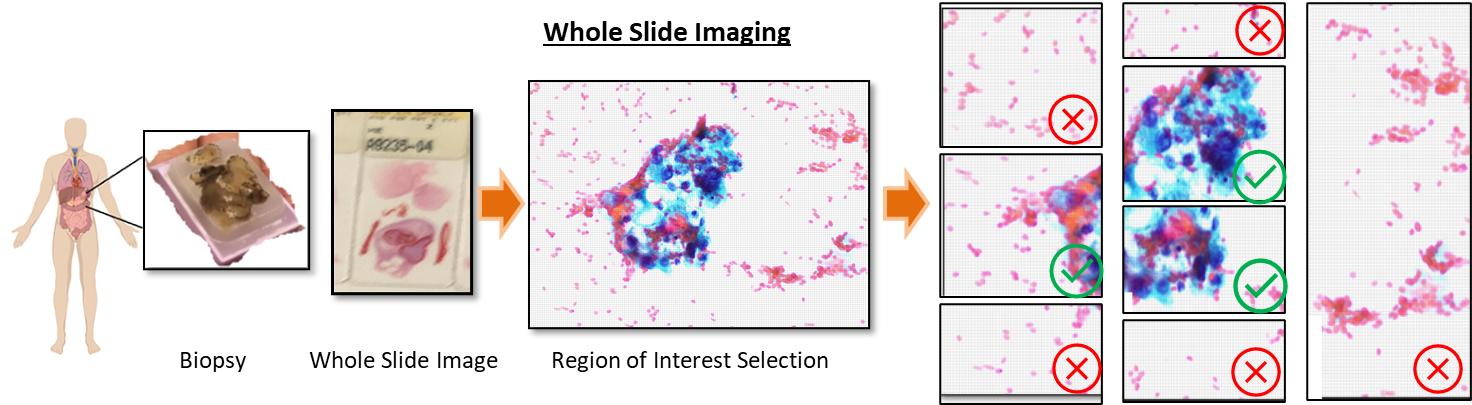}\label{fig:f1}}
  \caption{Whole Slide Image (WSI) processing and selection of tissue patches. Whole slide image was considered by dividing the slide into patches of 200x200 pixels. Each patch was then selected based on tissue presence.
}
\end{figure}

\subsection{Approximate Bayesian Neural Networks (BCNN) and Model Uncertainty}

During the training phase, histological pancreatic cancer image  dataset \(X=\left\{x_{1}, x_{2} \ldots x_{N}\right\}\) and the corresponding grade \(Y=\left\{y_{1}, y_{2} \ldots y_{N}\right\}\) where \(X \in R^{d}\) is a d-dimensional input vector and \(Y \in\{1 \ldots C\}\) with \(\mathrm{y}_{\mathrm{i}} \in\{1 \ldots \mathrm{C}\}\), given C class labels (here four: Normal, Grade I, Grade II and Grade III), a set of independent and identically distributed (i.i.d.) training samples size \(N\)\(\left\{x_{i}, y_{i}\right\}\)  for \(i=1\) to \(N\), are used to learn the weights of the neural networks. Uncertainty of the model prediction was captured by placing a prior distribution over weights \(W\). The principled predictive distribution of an unknown grading of cancer  prediction label \(\hat{y}\) of a test input data \(\hat{x}\) by marginalizing the parameters:

\begin{equation}
p(\hat{y}|\hat{x}, {X, Y}) = \int P(\hat{y} | \hat{x}, {w}) P({w}| X, Y, \hat{x}) d{w}
\end{equation}

Unfortunately, finding the posterior distribution \(p({w}|{X, Y})\) is often computationally intractable. Following Gal, Ghoshal et al. showed that neural networks with dropweights with the cross-entropy loss of the network, is equivalent to a variational approximation by minimising the Kullback-Leibler (KL) divergence on a Bayesian neural network. At test time, the unseen images are passed through the network, the posterior \(P({W}| X, Y, \hat{x})\) was approximated by averaging stochastic feed forward Monte Carlo (MC) sampling to estimate uncertainty.

Practically, the expectation of \(\hat{y}\) is called the predictive mean of the model. The estimate of the vector of Softmax probabilities i.e., the predictive mean \(\mu_{pred}\) over the MC iterations is then used as the final prediction on the test sample:
\begin{equation}
\mu_{pred} \approx \frac{1}{T} \sum_{t=1}^{T} p\left(\hat{y}=c | \hat{x}, \hat{\omega}_{\mathrm{t}}\right) ; \quad c \in\{1, \ldots, \mathrm{C}\}
\end{equation}

For each test sample \(\hat{x}\), the class with the largest predictive mean \(\mu_{pred}\) is selected as the output prediction. We present a novel approximation of predictive model uncertainty as below:

\begin{align}
    \text{$\sigma_{uncertainty}$ = }
    {
    \frac{1}{{C}} \sum_{{i}=1}^{{C}} \sqrt{\frac{1}{{T}} \sum_{{t}=1}^{{T}}\left[{p}\left(\hat{{y}}_{{t}}={c} | {\hat{x}}, \hat{\omega}_{{t}}\right)-{\hat\mu_{pred}}\right]^{2}}}
\end{align}
    \begin{equation}
    \text{where } \hat{y}_{t}=y\left(\hat{\omega}_{t}\right)=\text {Softmax}\left\{f^{\hat{\omega}_{t}} (\hat{x})\right\}
\label{eq:2}
\end{equation}

In our approximated uncertainty measure in model prediction (i.e. equation 3), we take into account the uncertainty associated with every class in the predictive mean $\mu_{pred}$. Furthermore in the approximation, we take the mean of the standard deviations of the class probabilities, instead of the variance. It assigns the highest average uncertainty to the most frequently mislabelled class.

\subsection{Uncertainty Metrics in Deep Learning}

We can compute an uncertainty metric from the multiple predictions per input image captured during test time. In this work we compared four well established metrics: predictive entropy; mutual information \cite{Gal,houlsby2011bayesian}, Feinman uncertainty, which measures of uncertainty for each observation by averaging the mean squared prediction error of each class \cite{feinman2017detecting}; Leibig uncertainty, which returns the empirical standard deviation as a proxy for predictive uncertainty \cite{leibig2017leveraging}; and moment based aleatoric and epistemic uncertainty metrics by Kwon \cite{Kwon}. In addition, we introduce a novel approximation of predictive model uncertainty which averages the standard deviations of the predictions in the predictive mean vector of class probabilities \cite{ghoshal2020calibrated,ghoshal2021cost}.

\subsection{Experiment}

Instead of training a very deep model from scratch on a small dataset, we decided
to run this experiment in a transfer learning setting, where we used a pretrained DenseNet-201, VGG-19 and ResNet-152V2 model \cite{Goodfellow} and acquired data only to fine-tune the original
model. We split the whole dataset into 60\% - 20\% - 20\% between training, validation and test sets respectively. This is suitable when the data is abound for an auxiliary domain,
but very limited labelled data is available for the domain of experiment. We introduced fully connected layers on top of the pre-trained convolutional base. Dropweights followed by a softmax activated layer is applied to the network as an approximation to the Gaussian Process (GP) and cast as an approximate Bayesian inference in the fully connected layer to estimate meaningful model uncertainty \cite{ghoshal2020estimating_covid}. The softmax layer outputs the probability distribution over each
possible class label.
We resized all images to 224 x 224 pixels (using a bicubic interpolation over
4 x 4 pixel neighbourhood). The images \cite{sehmi2021pancreatic} were then standardised using the mean and
standard deviation values of the MGG and H\&E dataset. 

Real-time data augmentation was also applied, leveraging Keras ImageDataGenerator during training, to prevent overfitting and enhance the learning capability of the model. Training images were
rotated 90 degrees, randomly flipped horizontally and vertically,scaled outward and inward, shifted, and sheared. The Adam optimiser was used with default initial learning rate of \(\alpha\) = 0.01 and moment decay rate of \(\beta1\) = 0.9 and \(\beta1\) = 0.999. All our experiments were
run for 100 epochs and batch size was set to 64. Dropweights with rates of 0.5 were added to the fully-connected layer. We monitored the validation accuracy after every epoch and saved the model with the best accuracy on the validation dataset. During test time, Dropweights were active and Monte Carlo
sampling was performed by feeding the input image with MC-samples 50 through the Bayesian Deep Convolutional Neural Networks.

\begin{figure}[!thp]
  \centering
  \subfloat {\includegraphics[width=1.0\textwidth]{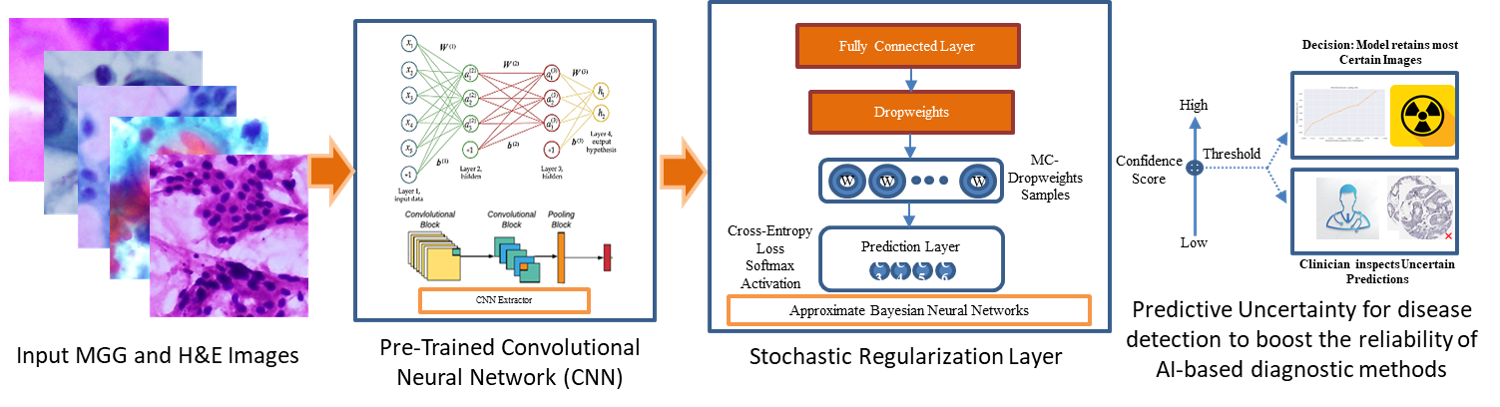}\label{fig:f1}}
  \caption{Overview of the proposed Bayesian Convolutional Neural Network (BCNN) framework.
}
\end{figure}

\section{Model Performance}

On average, Bayesian DenseNet-201 model based inference improves the detection
accuracy of the ResNet-152V2 and VGG-19 model in our sample dataset. Table 1 summarizes the Precision, Recall, F1-Score and prediction accuracy of our implemented models.

\begin{table}[!]
\resizebox{\textwidth}{!}{%
\begin{tabular}{|cc|cccc|cccc|cccc|}
\hline
\multicolumn{2}{|c|}{\textbf{Dataset}} & \multicolumn{4}{c|}{\textbf{VGG-19}} & \multicolumn{4}{c|}{\textbf{ResNet-152V2}} & \multicolumn{4}{c|}{\textbf{DenseNet-201}} \\ \hline
\multicolumn{2}{|c|}{\textbf{May Grunwald-Giemsa   (MGG)}} & \multicolumn{1}{c|}{\textbf{Precision}} & \multicolumn{1}{c|}{\textbf{Recall}} & \multicolumn{1}{c|}{\textbf{F1-Score}} & \textbf{Accuracy} & \multicolumn{1}{c|}{\textbf{Precision}} & \multicolumn{1}{c|}{\textbf{Recall}} & \multicolumn{1}{c|}{\textbf{F1-Score}} & \textbf{Accuracy} & \multicolumn{1}{c|}{\textbf{Precision}} & \multicolumn{1}{c|}{\textbf{Recall}} & \multicolumn{1}{c|}{\textbf{F1-Score}} & \textbf{Accuracy} \\ \hline
\multicolumn{2}{|c|}{} & \multicolumn{1}{c|}{} & \multicolumn{1}{c|}{} & \multicolumn{1}{c|}{} &  & \multicolumn{1}{c|}{} & \multicolumn{1}{c|}{} & \multicolumn{1}{c|}{} &  & \multicolumn{1}{c|}{} & \multicolumn{1}{c|}{} & \multicolumn{1}{c|}{} &  \\ \hline
\multicolumn{1}{|c|}{} & Normal & \multicolumn{1}{c|}{79.60\%} & \multicolumn{1}{c|}{78.80\%} & \multicolumn{1}{c|}{79.00\%} &  & \multicolumn{1}{c|}{80.20\%} & \multicolumn{1}{c|}{81.40\%} & \multicolumn{1}{c|}{80.60\%} &  & \multicolumn{1}{c|}{85.20\%} & \multicolumn{1}{c|}{83.40\%} & \multicolumn{1}{c|}{84.00\%} &  \\ \cline{2-14} 
\multicolumn{1}{|c|}{} & Grade I & \multicolumn{1}{c|}{53.00\%} & \multicolumn{1}{c|}{22.00\%} & \multicolumn{1}{c|}{31.50\%} &  & \multicolumn{1}{c|}{65.40\%} & \multicolumn{1}{c|}{40.80\%} & \multicolumn{1}{c|}{50.80\%} &  & \multicolumn{1}{c|}{72.20\%} & \multicolumn{1}{c|}{45.60\%} & \multicolumn{1}{c|}{55.40\%} &  \\ \cline{2-14} 
\multicolumn{1}{|c|}{} & Grade II & \multicolumn{1}{c|}{66.00\%} & \multicolumn{1}{c|}{82.40\%} & \multicolumn{1}{c|}{73.20\%} &  & \multicolumn{1}{c|}{75.80\%} & \multicolumn{1}{c|}{80.60\%} & \multicolumn{1}{c|}{78.20\%} &  & \multicolumn{1}{c|}{78.20\%} & \multicolumn{1}{c|}{84.20\%} & \multicolumn{1}{c|}{81.20\%} &  \\ \cline{2-14} 
\multicolumn{1}{|c|}{} & Grade III & \multicolumn{1}{c|}{50.20\%} & \multicolumn{1}{c|}{30.00\%} & \multicolumn{1}{c|}{38.00\%} &  & \multicolumn{1}{c|}{65.60\%} & \multicolumn{1}{c|}{58.60\%} & \multicolumn{1}{c|}{61.20\%} &  & \multicolumn{1}{c|}{72.20\%} & \multicolumn{1}{c|}{66.00\%} & \multicolumn{1}{c|}{68.80\%} &  \\ \hline
\multicolumn{2}{|c|}{Average Score} & \multicolumn{1}{c|}{62.89\%} & \multicolumn{1}{c|}{66.69\%} & \multicolumn{1}{c|}{63.66\%} & 66.69\% & \multicolumn{1}{c|}{74.14\%} & \multicolumn{1}{c|}{74.12\%} & \multicolumn{1}{c|}{73.77\%} & 74.23\% & \multicolumn{1}{c|}{78.19\%} & \multicolumn{1}{c|}{78.19\%} & \multicolumn{1}{c|}{77.86\%} & 85.53\% \\ \hline
\multicolumn{2}{|c|}{\textbf{}} & \multicolumn{1}{c|}{} & \multicolumn{1}{c|}{} & \multicolumn{1}{c|}{} &  & \multicolumn{1}{c|}{} & \multicolumn{1}{c|}{} & \multicolumn{1}{c|}{} &  & \multicolumn{1}{c|}{} & \multicolumn{1}{c|}{} & \multicolumn{1}{c|}{} &  \\ \hline
\multicolumn{2}{|c|}{\textbf{Haematoxylin and   Eosin (H\&E)}} & \multicolumn{1}{c|}{} & \multicolumn{1}{c|}{} & \multicolumn{1}{c|}{} &  & \multicolumn{1}{c|}{} & \multicolumn{1}{c|}{} & \multicolumn{1}{c|}{} &  & \multicolumn{1}{c|}{} & \multicolumn{1}{c|}{} & \multicolumn{1}{c|}{} &  \\ \hline
\multicolumn{2}{|c|}{} & \multicolumn{1}{c|}{} & \multicolumn{1}{c|}{} & \multicolumn{1}{c|}{} &  & \multicolumn{1}{c|}{} & \multicolumn{1}{c|}{} & \multicolumn{1}{c|}{} &  & \multicolumn{1}{c|}{} & \multicolumn{1}{c|}{} & \multicolumn{1}{c|}{} &  \\ \hline
\multicolumn{1}{|c|}{} & Normal & \multicolumn{1}{c|}{87.60\%} & \multicolumn{1}{c|}{83.40\%} & \multicolumn{1}{c|}{85.20\%} &  & \multicolumn{1}{c|}{90.00\%} & \multicolumn{1}{c|}{87.80\%} & \multicolumn{1}{c|}{88.80\%} &  & \multicolumn{1}{c|}{90.00\%} & \multicolumn{1}{c|}{90.00\%} & \multicolumn{1}{c|}{90.00\%} &  \\ \cline{2-14} 
\multicolumn{1}{|c|}{} & Grade I & \multicolumn{1}{c|}{84.40\%} & \multicolumn{1}{c|}{79.60\%} & \multicolumn{1}{c|}{81.80\%} &  & \multicolumn{1}{c|}{85.40\%} & \multicolumn{1}{c|}{86.80\%} & \multicolumn{1}{c|}{86.20\%} &  & \multicolumn{1}{c|}{88.20\%} & \multicolumn{1}{c|}{88.40\%} & \multicolumn{1}{c|}{88.20\%} &  \\ \cline{2-14} 
\multicolumn{1}{|c|}{} & Grade II & \multicolumn{1}{c|}{62.40\%} & \multicolumn{1}{c|}{77.40\%} & \multicolumn{1}{c|}{68.80\%} &  & \multicolumn{1}{c|}{79.60\%} & \multicolumn{1}{c|}{79.60\%} & \multicolumn{1}{c|}{79.80\%} &  & \multicolumn{1}{c|}{81.00\%} & \multicolumn{1}{c|}{81.40\%} & \multicolumn{1}{c|}{81.20\%} &  \\ \cline{2-14} 
\multicolumn{1}{|c|}{} & Grade III & \multicolumn{1}{c|}{76.80\%} & \multicolumn{1}{c|}{65.80\%} & \multicolumn{1}{c|}{70.80\%} &  & \multicolumn{1}{c|}{79.80\%} & \multicolumn{1}{c|}{78.60\%} & \multicolumn{1}{c|}{79.80\%} &  & \multicolumn{1}{c|}{83.20\%} & \multicolumn{1}{c|}{82.60\%} & \multicolumn{1}{c|}{82.80\%} &  \\ \hline
\multicolumn{2}{|c|}{Average Score} & \multicolumn{1}{c|}{78.02\%} & \multicolumn{1}{c|}{76.51\%} & \multicolumn{1}{c|}{76.79\%} & 76.52\% & \multicolumn{1}{c|}{83.34\%} & \multicolumn{1}{c|}{83.48\%} & \multicolumn{1}{c|}{83.62\%} & 83.60\% & \multicolumn{1}{c|}{85.65\%} & \multicolumn{1}{c|}{85.71\%} & \multicolumn{1}{c|}{85.61\%} & { \textbf{85.60\%}} \\ \hline
\multicolumn{2}{|c|}{\textbf{}} & \multicolumn{1}{c|}{} & \multicolumn{1}{c|}{} & \multicolumn{1}{c|}{} &  & \multicolumn{1}{c|}{} & \multicolumn{1}{c|}{} & \multicolumn{1}{c|}{} &  & \multicolumn{1}{c|}{} & \multicolumn{1}{c|}{} & \multicolumn{1}{c|}{} &  \\ \hline
\multicolumn{2}{|l|}{\textbf{Mixed (H\&E and MGG)}} & \multicolumn{1}{c|}{} & \multicolumn{1}{c|}{} & \multicolumn{1}{c|}{} &  & \multicolumn{1}{c|}{} & \multicolumn{1}{c|}{} & \multicolumn{1}{c|}{} &  & \multicolumn{1}{c|}{} & \multicolumn{1}{c|}{} & \multicolumn{1}{c|}{} &  \\ \hline
\multicolumn{2}{|c|}{} & \multicolumn{1}{c|}{} & \multicolumn{1}{c|}{} & \multicolumn{1}{c|}{} &  & \multicolumn{1}{c|}{} & \multicolumn{1}{c|}{} & \multicolumn{1}{c|}{} &  & \multicolumn{1}{c|}{} & \multicolumn{1}{c|}{} & \multicolumn{1}{c|}{} &  \\ \hline
\multicolumn{1}{|c|}{} & Normal & \multicolumn{1}{c|}{66.40\%} & \multicolumn{1}{c|}{57.00\%} & \multicolumn{1}{c|}{59.00\%} &  & \multicolumn{1}{c|}{79.00\%} & \multicolumn{1}{c|}{82.60\%} & \multicolumn{1}{c|}{80.60\%} &  & \multicolumn{1}{c|}{82.20\%} & \multicolumn{1}{c|}{86.20\%} & \multicolumn{1}{c|}{84.20\%} &  \\ \cline{2-14} 
\multicolumn{1}{|c|}{} & Grade I & \multicolumn{1}{c|}{79.60\%} & \multicolumn{1}{c|}{66.80\%} & \multicolumn{1}{c|}{72.60\%} &  & \multicolumn{1}{c|}{84.00\%} & \multicolumn{1}{c|}{80.00\%} & \multicolumn{1}{c|}{81.80\%} &  & \multicolumn{1}{c|}{85.80\%} & \multicolumn{1}{c|}{78.60\%} & \multicolumn{1}{c|}{82.00\%} &  \\ \cline{2-14} 
\multicolumn{1}{|c|}{} & Grade II & \multicolumn{1}{c|}{57.60\%} & \multicolumn{1}{c|}{83.00\%} & \multicolumn{1}{c|}{63.80\%} &  & \multicolumn{1}{c|}{75.00\%} & \multicolumn{1}{c|}{80.40\%} & \multicolumn{1}{c|}{77.80\%} &  & \multicolumn{1}{c|}{76.80\%} & \multicolumn{1}{c|}{81.80\%} & \multicolumn{1}{c|}{79.20\%} &  \\ \cline{2-14} 
\multicolumn{1}{|c|}{} & Grade III & \multicolumn{1}{c|}{40.80\%} & \multicolumn{1}{c|}{37.00\%} & \multicolumn{1}{c|}{54.50\%} &  & \multicolumn{1}{c|}{72.60\%} & \multicolumn{1}{c|}{63.20\%} & \multicolumn{1}{c|}{67.60\%} &  & \multicolumn{1}{c|}{75.00\%} & \multicolumn{1}{c|}{69.60\%} & \multicolumn{1}{c|}{71.80\%} &  \\ \hline
\multicolumn{2}{|c|}{\textbf{Average Score}} & \multicolumn{1}{c|}{60.55\%} & \multicolumn{1}{c|}{59.81\%} & \multicolumn{1}{c|}{55.12\%} & {59.75\%} & \multicolumn{1}{c|}{77.19\%} & \multicolumn{1}{c|}{77.16\%} & \multicolumn{1}{c|}{77.08\%} & 77.16\% & \multicolumn{1}{c|}{79.35\%} & \multicolumn{1}{c|}{79.33\%} & \multicolumn{1}{c|}{79.15\%} & 79.29\% \\ \hline
\end{tabular}
}
\caption{summarizes the Precision, Recall, F1-Score and prediction accuracy of our implemented models}
\end{table}

All models trained with the H\&E stained images achieved the highest accuracy and F1-score compared to MGG and mixed. VGG-19 model trained with the mixed dataset performed the lowest compared to H\&E and MGG which indicates H\&E stained images are easier to learn and achieve better prediction than MGG. All models achieved an exceptional performance in identifying normal tissue from H\&E stained images, which could be due to the distinct differences between neoplastic and non-neoplastic tissue. We also observed that all models performed the weakest in Grade-III class prediction compared other grades.

\section{Prediction Error vs Uncertainty}
\subsection{Bayesian Model Uncertainty}

We measured the uncertainty associated with the predictive probabilities of the deep learning model by keeping dropweights on during test time. Figure 3 shows that the pancreatic adenocarcinoma grade-specific approximations of uncertainty is consistent with the confusion matrix. Higher uncertainty seems to be associated with slides that tend to be misclassified.

\begin{figure}[!thp]
  \centering
 {\includegraphics[width=1.0\textwidth]{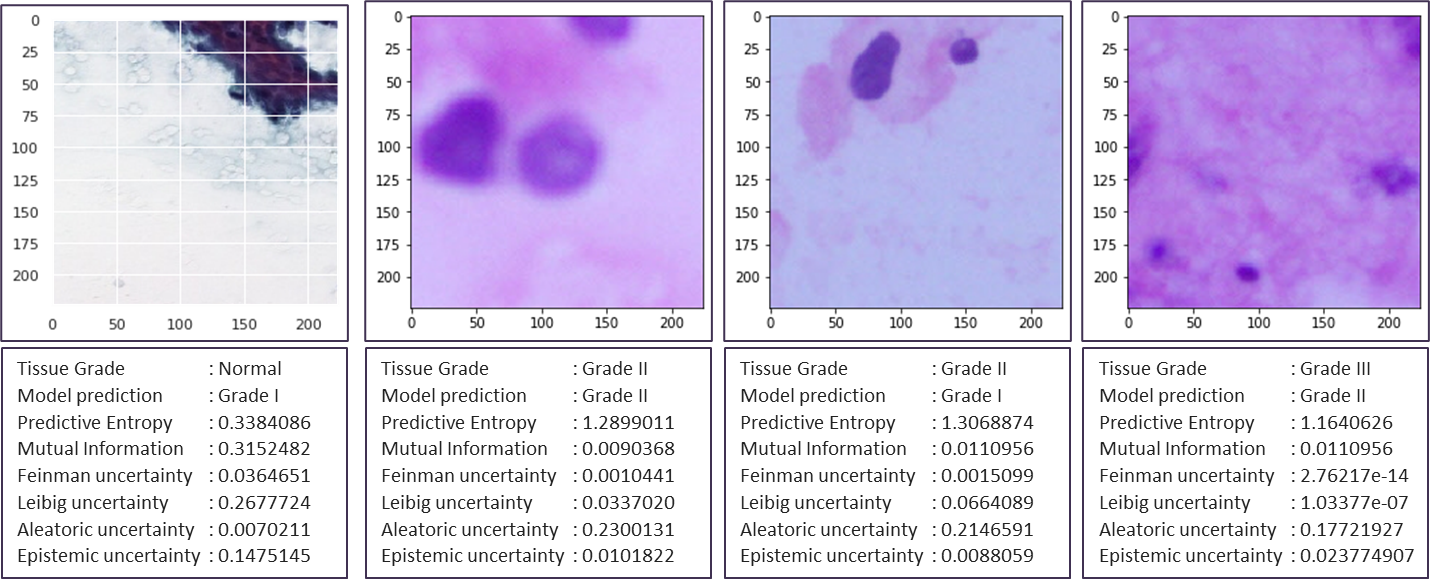}\label{fig:f1}}
  {\includegraphics[width=1.0\textwidth]{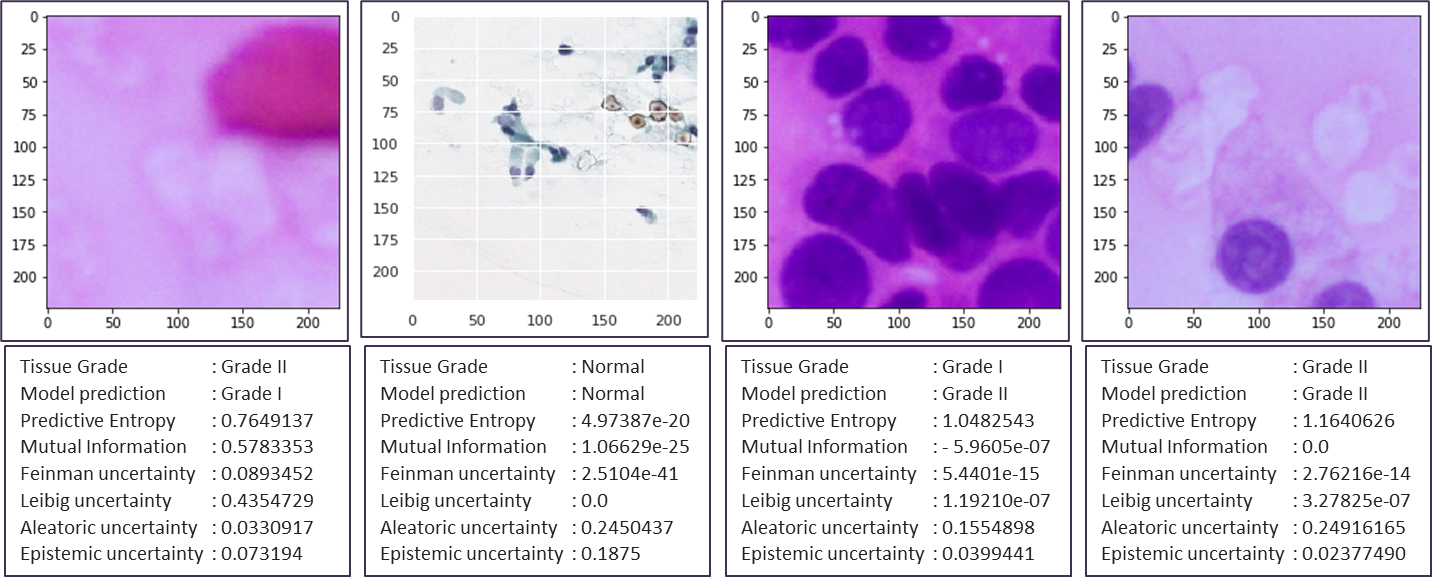}\label{fig:f2}}
  \caption{Example Test Images with Estimated Uncertainty }
\end{figure}

Figure 4 shows the boxplots of estimated uncertainty in deep learning model for pancreatic adenocarcinoma grading. 

\begin{figure}[!thp]
  \centering
 {\includegraphics[width=1.0\textwidth]{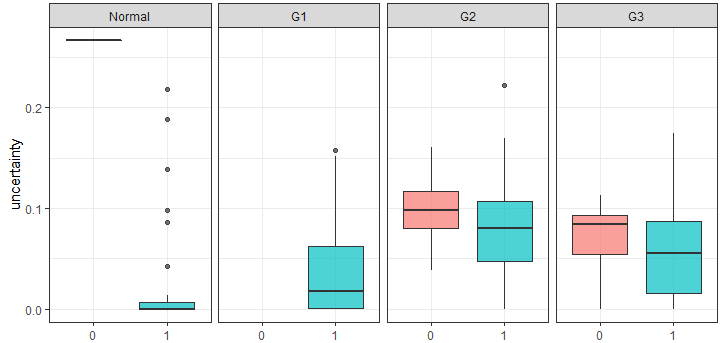}\label{fig:f1}}
  \caption{The red box indicates the uncertainty of incorrectly graded images. The
green box corresponds to correctly classified images. Our model assigns the highest average uncertainty to the mislabelled grades and less contribution to high uncertainty values by correctly classified images. Therefore, uncertainty information provides as an additional insight to point prediction to refer the uncertain images to radiologists for further investigation, which improves the overall prediction performance. }
\end{figure}

\subsection{Relationship between the Accuracy and Uncertainty}

Trustworthy clinical Artificial Intelligence (AI) systems should not only return accurate predictions, but also a credible representation of their uncertainty. The model’s expected accuracy increases as the model’s uncertainty in prediction decreases \cite{ghoshal2021uncertaintySAR}.

The true error is the difference between estimated values and actual values. In order to assess the quality of predictive uncertainty, we leveraged Spearman’s correlation coefficient. We quantified the predictive accuracy by 1-Wasserstein distance (WD) to measure how much the estimated uncertainty correlates with the true errors \cite{arjovsky2017wasserstein,laves2019quantifying}.
The Wasserstein distance for the real data distribution \(P_r\) and the generated data distribution \(P_g\) is mathematically defined as the greatest lower bound (infimum) for any transport plan (i.e. the cost for the cheapest plan):
\begin{equation}
W(P_r, P_g) = \inf_{\gamma \sim \Pi(P_r, P_g)} \mathbb{E}_{(x, y) \sim \gamma}[\| x-y \|]
\end{equation}, \(\Pi(P_r, P_g)\) is the set of all possible joint probability distributions \(\gamma(x, y)\) whose marginals are respectively \(P_r\) and \(P_g\). However, the equation (5) for the Wasserstein distance is intractable. Using the Kantorovich-Rubinstein duality it can be simplified to

\begin{equation}
W(P_r, P_g) = \frac{1}{K} \sup_{\| f \|_L \leq K} \mathbb{E}_{x \sim P_r}[f(x)] - \mathbb{E}_{x \sim P_g}[f(x)]
\end{equation}, where sup (supremum) is the least upper bound and \(f\) is a 1-Lipschitz continuous functions \(\{ f_w \}_{w \in W}\), parameterized by \(w\) and the K-Lipschitz constraint \(\lvert f(x_1) - f(x_2) \rvert \leq K \lvert x_1 - x_2 \rvert\). The error function can be configured as measuring the 1 - Wasserstein distance between \(P_r\) and \(P_g\).
\begin{equation}
E(P_r, P_g) = W(P_r, P_g) = \max_{w \in W} \mathbb{E}_{x \sim P_r}[f_w(x)] - \mathbb{E}_{z \sim P_r(z)}[f_w(g_\theta(z))]
\end{equation}

The advantage of Wasserstein distance (WD) is that it can reflect the distance of two non-overlapping or little overlapping distributions.


The figure 5 below shows the correlation between estimated uncertainty and the error of prediction and Spearman correlation. The results show strong correlation with \(\rho\) > 0.95 between entropy of the probabilities as a measure of the epistemic uncertainty and prediction errors.

\begin{figure}[!thp]
  \centering
  \subfloat[DenseNet-201] {\includegraphics[width=0.33\textwidth]{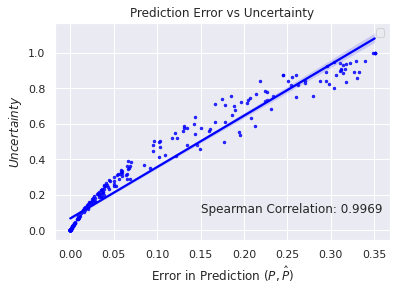}\label{fig:f1}}
  \subfloat[ResNet-152V2] {\includegraphics[width=0.33\textwidth]{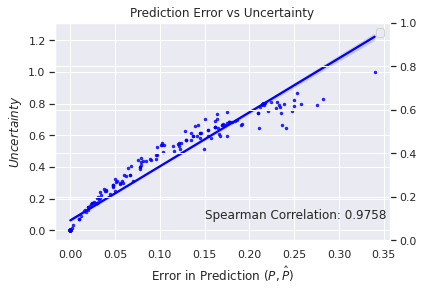}\label{fig:f2}}
  \subfloat[VGG-19] {\includegraphics[width=0.33\textwidth]{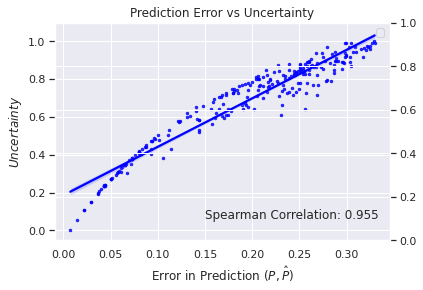}\label{fig:f2}}
  \caption{Correlation between estimated predictive entropy as a measure of Uncertainty and Accuracy in prediction}
\end{figure}

Our experiments show that the prediction uncertainty correlates with accuracy, thus enabling the  identification of false predictions or unknown cases.

\subsection{Performance improvement via Uncertainty-Aware Cancer grading Classification}

We performed predictions for all test images and sorted the predictions by their associated predictive uncertainty. We then referred predictions based on the various levels of  uncertainty for further diagnosis and measured the accuracy of the predictions for the remaining cases. We observed in figure 6, uncertainty estimation can usually be used in every image classifier to improve prediction accuracy of man–machine combination via uncertainty-aware referral with the additional computational load cost of performing multiple forward passes.

\begin{figure}[!thp]
  \centering
  \subfloat[DenseNet-201] {\includegraphics[width=0.33\textwidth]{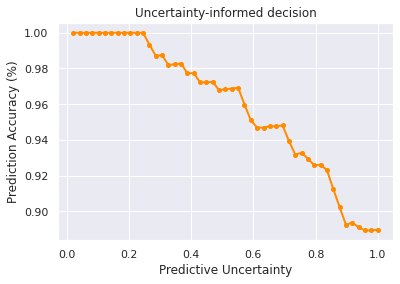}\label{fig:f1}}
  \subfloat[VGG-19] {\includegraphics[width=0.33\textwidth]{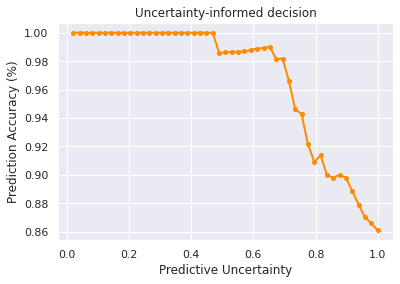}\label{fig:f2}}
  \subfloat[ResNet-152V2] {\includegraphics[width=0.33\textwidth]{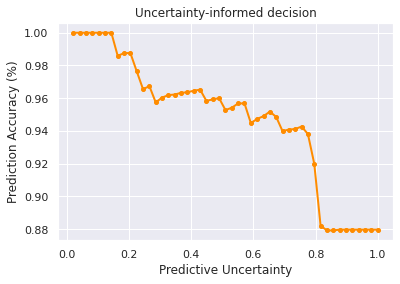}\label{fig:f2}}
  \caption{The classification accuracy as a function of the tolerated normalized model
  uncertainty}
\end{figure}

\section{Leveraging uncertainty in classification error and reject tradeoff}

In safety critical decision making scenarios, the goal of a deep learning model is that the algorithm abstains from making a prediction on the most uncertain images rather than making an incorrect prediction, where the risk of making a incorrect prediction is too large i.e. allowing the model to say "I don't know".

A low model uncertainty ascertained when the probability of predicting into the most possible category has much greater margin over the 2nd most likely category \cite{ghoshal2021deephistoclass}. In rejection, the prediction was ambiguous among all categories because the model failed to reach a definitive conclusion between all the grades.

For example, if the model was presented with a out-of-distribution or non-domain image, it would still classify the image into one of the 4 grades albeit with a high estimated uncertainty measure, which is not an expected outcome. Instead, the model should be able to reject predicting the grade for the test images where the uncertainty is too high. Such abstention is known as classification with a reject option. In the context of the clinical diagnostic process, accuracy is of paramount importance. The overall decision is the following: the model would accept if the acceptance criteria defined as $\text{$\Lambda(x)$}$ defined as \cite{chow1970optimum,Sebastian}: 
\begin{equation}
    \text{$\Lambda(x)$} :=
	\vert P(\mathbf{x}^{(i)})_1 - P(\mathbf{x}^{(i)})_2 \vert \geq \epsilon * [ {\textrm{$\sigma_{uncertainty}$}}(\mathbf{x}^{(i)})_1]
\end{equation}
where $P(\mathbf{x}^{(i)})$ is the predictive probability for input image $\mathbf{x}^{(i)}$ with $P(\mathbf{x}^{(i)})_j$ being defined as the probability of the $j\textsuperscript{th}$ most likely class, and ${\textrm{U}}(\mathbf{x}^{(i)})_1$ is the estimated uncertainty of the most likely class returned by the model for input $\mathbf{x}^{(i)}$. Note that the $\epsilon$ probability threshold parameter defines a trade-off between the number of  classified examples at the number of examples that would have been incorrectly classified but were accepted instead.

\begin{figure}[!thp]
  \centering
  \subfloat {\includegraphics[width=0.8\textwidth]{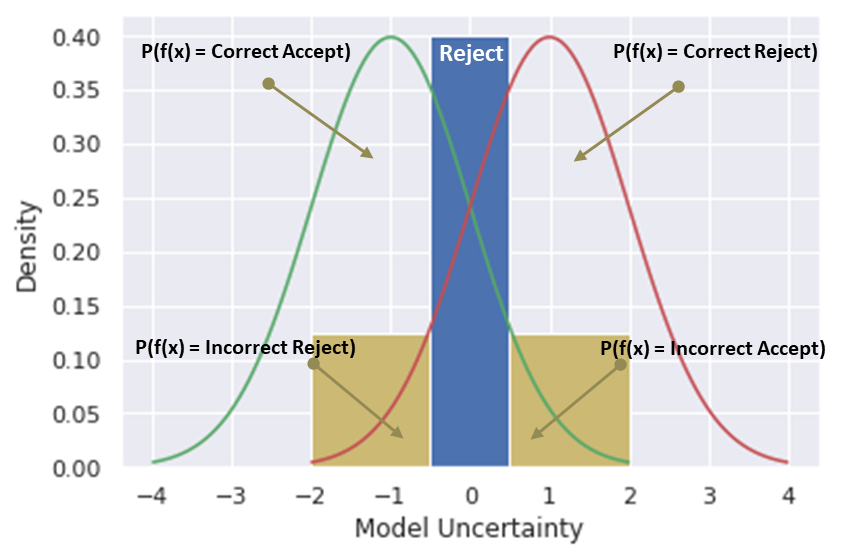}\label{fig:f1}}
  \caption{Distribution of predictive uncertainty values grouped by correct and erroneous predictions
}
\end{figure}

Figure 7 illustrates two class-conditional distribution of predictive uncertainty for the discriminant. The conditional accept or reject with estimated uncertainty depends on uncertainty thresholds.

\subsection{Evaluation Metric - Accuracy‐Rejection Quotient (ARQ)}

The standard metric for evaluating a model $f$ on a classification task is the accuracy: 
\begin{equation}
	\textrm{Accuracy} = \frac{1}{N} \sum_{i=1}^N \mathbbm{1}(P(\mathbf{x}^{(i)}) = y^{(i)})
\end{equation}

where $N$ is the number of test data points and $(\mathbf{x}^{(i)}, y^{(i)})$ is the $i\textsuperscript{th}$ test point. Note that the accuracy does not take into account the uncertainty with which a model makes the prediction and only rewards correctly classified examples.

The metric Accuracy‐Rejection Quotient (ARQ)  for classification with a accuracy-reject trade-off by assigning a cost $\alpha$ to misclassification, and a cost $\beta$ to the acceptance option $\text{$\Lambda(x)$}$ is  defined as \cite{chow1970optimum,Sebastian}:
\begin{equation}
	\textrm{ARQ}_{\alpha, \beta} = \frac{1}{N} \sum_{i=1}^N \mathbbm{1}(P(\mathbf{x}^{(i)}) = y^{(i)}) - \alpha \mathbbm{1}(P(\mathbf{x}^{(i)}) \neq y^{(i)}) - \beta \mathbbm{1}(P(\mathbf{x}^{(i)}) \neq \text{$\Lambda(\mathbf{x}^{(i)})$} )
\end{equation}
Here, $\alpha$ and $\beta$ are hyperparameters. A large value for $\alpha$ corresponds to cases where cost of misclassification is extremely high such as in medical diagnosis. The $\beta$ intuitively is a trade-off indicator - higher $\beta$ will decrease the cost of the accept option.

The figure 8 shows the trade-off between the number of misclassified but accepted images and the number of correctly classified but rejected images.

\begin{figure}[!thp]
  \centering
  \subfloat[DenseNet-201] {\includegraphics[width=0.33\textwidth]{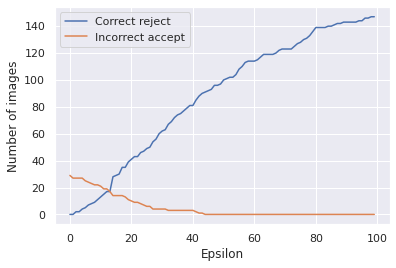}\label{fig:f1}}
  \subfloat[VGG-19] {\includegraphics[width=0.33\textwidth]{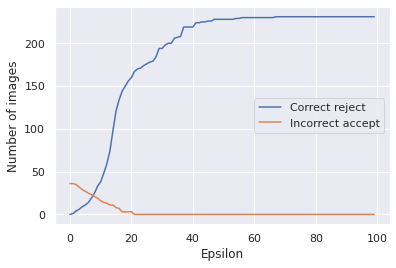}\label{fig:f2}}
  \subfloat[ResNet-152V2] {\includegraphics[width=0.33\textwidth]{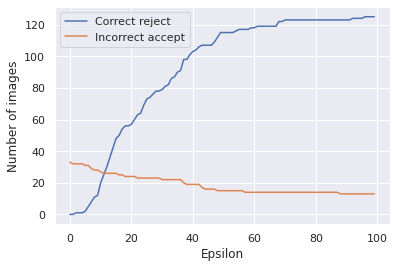}\label{fig:f2}}
  \caption{Accuracy‐Rejection Quotient (ARQ) obtained using varying $\epsilon$ and $\alpha$ but setting $\beta=0$. For $\alpha>0$ there seems to be an optimal value of $\epsilon$ around 17 in case of DenseNet-201.
}
\end{figure}

\section{Conclusion and Future work}

Clinical AI systems are now regularly being used in medical settings although regulatory oversight is inconsistent and undeveloped. Computer based medical systems in clinical setting requires informed users, who are generally responsible for identifying and reporting emerging problems \cite{rajpurkar2022ai}. Understanding the confidence of predictions made by a deep learning model can aide clinicians identify when AI systems fail is essential for gaining clinicians' trust in the technology. In this work, Bayesian Deep Learning classifier has been trained using transfer
learning method on pancreatic cancer grade from histopathology images to estimate model uncertainty. The evaluation shows that the Bayesian DenseNet-201 model trained by H\&E stained image dataset achieved the best performance compared to the other models and dataset. Our experiment has shown a strong correlation between model uncertainty and error in prediction. We have also shown how to leverage estimated uncertainty in prediction as rejection threshold in classifying images by user-defined hyperparameters for a given cost of misclassification and rejection cost to control the accuracy-acceptance rate of the model. The estimated uncertainty in deep learning yields more reliable predictions; protects against model limitations from possible scenarios in training dataset such as minority class, dismissal, automation bias or  out-of-distribution test dataset, which can alert radiologists on false predictions, increasing the acceptance of deep learning into clinical practice in disease detection. With this Bayesian Deep Learning based classification, studies correlating with multi "omics" datasets ( pathology with radiological, genomic, and proteomic) and treatment responses could further reveal insights about imaging markers and findings towards improved diagnosis and treatment for pancreatic adenocarcinomas.

\bibliographystyle{splncs04}
\bibliography{MIUA_Final_Submission_V1.0}
\end{document}